\documentclass[preprint,prd,aps,showpacs,preprintnumbers,amsmath,amssymb]{revtex4-1}
\usepackage{graphics,epsfig,subfigure}
\usepackage{diagbox}
\usepackage{CJKutf8}
\usepackage[usenames]{color}
\usepackage[colorlinks,
            linkcolor=black,
            anchorcolor=blue,
            citecolor=blue
            ]{hyperref}
\usepackage{color}
\usepackage{graphicx}
\usepackage{amsfonts}
\usepackage{indentfirst}
\usepackage{booktabs}

\newcommand\beq{\begin{equation}}
\newcommand\eeq{\end{equation}}
\newcommand\beqn{\begin{eqnarray}}
\newcommand\eeqn{\end{eqnarray}}

\begin{document}
\begin{CJK*}{UTF8}{gbsn}

\title{\Large \bf $\kappa$-Dirac stars}
\author{Shi-Xian Sun, Si-Yuan Cui, Long-Xing Huang and Yong-Qiang Wang\footnote{yqwang@lzu.edu.cn, corresponding author}}

\affiliation{$^{1}$Institute of Theoretical Physics and Research Center of Gravitation, Lanzhou University, Lanzhou 730000, China\\
	$^{2}$Key Laboratory of Quantum Theory and Applications of MoE, Lanzhou University, Lanzhou 730000, China\\
    $^{3}$Lanzhou Center for Theoretical Physics and Key Laboratory of Theoretical Physics of Gansu Province, Lanzhou University, Lanzhou 730000, China}

\begin{abstract}
In this paper, we construct a Dirac star model composed of $|\kappa|$ pairs of spinor fields. The azimuthal harmonic indeces $m$ of these spinor fields are half-integers, and they satisfiy  $-(|\kappa|-\frac{1}{2})\leq m \leq |\kappa|-\frac{1}{2}$. When $\kappa=1$, it corresponds to the conventional Dirac star model, formed by two spinor fields with $m=\frac{1}{2}$ and $m=-\frac{1}{2}$. When $|\kappa|>1$, among these $|\kappa|$ pairs of spinor fields, there exist spinor fields with azimuthal harmonic indices $m>\frac{1}{2}$, and all spinor fields still conform to the same Dirac field equation. Different families of solutions are distinguished by the value of $\kappa$, so we named these solutions $\kappa$-Dirac stars. We obtain solutions for $\kappa=\pm1,\pm2,\pm3,\pm4,\pm5,\pm6$ by using numerical methods. Additionally, we compute their ADM mass $M$, Noether charge $Q$, and binding energy $E$, and illustrate how these quantities change with the spinor field's frequency $\omega$ for different $\kappa$. We observe significant differences between solutions for $|\kappa|>1$ and the $|\kappa|=1$ case. Furthermore, we provide the energy density distribution of the Dirac stars, wherein for $|\kappa|>1$ scenarios, the Dirac stars exhibit a spherical shell-like structure. Moreover, we employ three-dimensional diagrams to intuitively depict how $\kappa$ influences the combination of spinor fields to form a spherically symmetric configuration.
\end{abstract}

\maketitle

\section{Introduction}\label{Sec1}

The study of particle-like solutions has a long and extensive history. In the context of general relativity, a crucial question arises: can classical fields, under the influence of their own gravity, form energy-localized configurations without singularities? This query was first explored by Wheeler in 1955, who envisioned that electromagnetic waves could give rise to such particle-like solutions through their self-gravity, named ``geons" \cite{Wheeler:1955zz}. In the 1960s, Kaup \cite{Kaup:1968zz} and Ruffini \cite{Ruffini:1969qy} introduced massive complex scalar fields as substitutes for the electromagnetic field, resulting in stable localized solutions known as Klein-Gordon Geon, later referred to as boson stars. Not only scalar fields but also massive complex vector fields can exhibit similar configurations, known as Proca stars \cite{Rosen:1994rq,Brito:2015pxa}. Bosonic stars find applications in various astrophysical phenomena such as candiates for dark matter \cite{Sahni:1999qe,Padilla:2019fju}, sources of gravitational waves \cite{CalderonBustillo:2020fyi,Zhang:2021ojz,Bezares:2022obu}, and black hole mimickers \cite{Cardoso:2019rvt,Herdeiro:2021lwl,Rosa:2023qcv,Rosa:2022tfv}. In addition to serving as constituents of horizonless exotic compact objects, the study of scalar fields and Proca fields as black hole hair is also highly extensive\cite{Herdeiro:2014goa,Aguilar-Nieto:2022jio,Teodoro:2021ezj,Cunha:2015yba,East:2017ovw,Wang:2018xhw}.

Furthermore, considering spin-$1/2$ spinor fields as classical fields coupled to gravity can also yield particle-like solutions \cite{Finster:1998ws}. In Einstein-Dirac system, such solutions can be called ``Dirac stars". However, unlike the study of boson stars and Proca stars, obtaining spherically symmetric solutions for a single spinor field is not possible due to its intrinsic angular momentum. Thus, at least two spinor fields with opposite angular momentum directions are required to form a spherically symmetric spacetime. Similar construction methods have also been applied in the study of cosmology \cite{Armendariz-Picon:2003wfx,Finster:2009lcp} and wormholes \cite{Blazquez-Salcedo:2020czn,Bolokhov:2021fil,Konoplya:2021hsm,Wang:2022aze,Kain:2023ann}.

Just like the study of boson stars can be extended \cite{Colpi:1986ye,Jetzer:1989av,Yoshida:1997qf,Kleihaus:2009kr,Collodel:2022jly,Jaramillo:2022gcq,Ogawa:2023ive}, the study of Dirac stars can also be generalized to include electric charge \cite{Finster:1998ux,Herdeiro:2021jgc}, self-interactions \cite{Dzhunushaliev:2018jhj,Herdeiro:2020jzx}, higher dimensions \cite{Blazquez-Salcedo:2019qrz}, excited states \cite{Herdeiro:2017fhv} and hybrid stars \cite{Liang:2022mjo,Liang:2023ywv}. In the case of rotating Dirac stars, spherical symmetry is not required, and they can be constructed using a single spinor field, yielding solutions with azimuthal harmonic indices $m>1/2$ \cite{Herdeiro:2019mbz,Huang:2023glq}. However, the primary focus of research on spherically symmetric Dirac stars has centered around those formed by two spinor fields with azimuthal harmonic indices $m=1/2$. In the context of spherical symmetry, multiple spinor fields with higher total angular momentum can also give rise to solitonic solutions \cite{Finster:1998ju,Leith:2020jqw}. Furthermore, these solutions also can exhibit excited states \cite{Leith:2021urf}. This model also admits solutions where the spinor fields possess negative eigenvalues. The solutions with negative eigenvalues was mentioned in some studies \cite{Finster:1998ju,Blazquez-Salcedo:2020czn,Leith:2021urf}, but had not undergone a more in-depth investigation. This paper thoroughly studies spherically symmetric Dirac stars composed of multiple spinor fields, incorporating scenarios where the spinor fields possess negative eigenvalues.

The organization of this paper is as follows. In Sec. \ref{Sec2}, we present the $\kappa$-Dirac star model, which couples $|\kappa|$ pairs of spinor fields with Einstein gravity. In Sec. \ref{Sec3}, we outline the boundary conditions that the gravitational field and spinor fields must satisfy. The Sec. \ref{Sec4} showcases numerical results, highlighting the distinctions between Dirac stars formed by a single pair of spinor fields and those formed by multiple pairs of spinor fields. In the final section, we make a summary and offer prospects for future research.

\section{The Model Setup}\label{Sec2}

We consider a system composed of multiple spinor fields minimally coupled to (3 + 1)-dimensional Einstein's gravity. The action is given by
\begin{equation}
\label{L}
\mathcal{S}=\int d^4 x \sqrt{-g}\left[\frac{R}{16 \pi G}+\mathcal{L}_{D}\right],
\end{equation}
where $R$ is the Ricci scalar, $G$ is the gravitational constant, second term $\mathcal{L}_{D}$  represents the Lagrangian density of multiple spinor fields, expressed as:
\begin{equation}
\label{LD}
\mathcal{L}_D = -i\sum_{j=1}^{|2\kappa|} \left[\frac{1}{2}(\hat{D}_\mu \bar{\Psi}^{(j)} \gamma^\mu \Psi^{(j)}-\bar{\Psi}^{(j)} \gamma^\mu \hat{D}_\mu \Psi^{(j)})+\mu \bar{\Psi}^{(j)} \Psi^{(j)}\right].
\end{equation}
Here, ${\Psi}^{(j)}$ ($j = 1,2,...,|2\kappa|$, $\kappa$ is a non-zero integer) are the spinor fields. $\bar{\Psi}^{(j)}=\Psi^{(j)\dagger}\xi$ are the Dirac conjugate of ${\Psi}^{(j)}$, with $\Psi^{(j)\dagger}$ denoting the usual Hermitian conjugate. For the Hermitian matrix $\xi$, we choose $\xi=-\hat{\gamma}^0$, where $\hat{\gamma}^0$ is one of gamma matrices in flat spacetime. $\hat{D}_\mu=\partial_\mu-\Gamma_\mu$, where $\Gamma_\mu$ are the spinor connection matrices. All spinor fields have the same mass $\mu$. We present further details of the spinor fields in Appendix A. By varying the action, we can obtain the Einstein equation and the Dirac equation:
\begin{equation}
\label{Eequ}
E_{\mu\nu}\equiv R_{\mu\nu}-\frac{1}{2} g_{\mu\nu}R-8\pi G T_{\mu\nu}=0,
\end{equation}
\begin{equation}
\label{Dequ}
\gamma^\mu\hat{D_\mu}\Psi^{(j)}-\mu\Psi^{(j)}=0,
\end{equation}
where, $T_{\mu\nu}$ is the total energy-momentum tensor, obtained by summing the energy-momentum tensors of each individual spinor field,
\begin{equation}
\label{T}
T_{\mu\nu}=\sum_{j=1}^{|2\kappa|}\frac{i}{2}(\hat{D}_\mu \bar{\Psi}^{(j)} \gamma_\nu \Psi^{(j)}+\hat{D}_\nu \bar{\Psi}^{(j)} \gamma_\mu \Psi^{(j)}-\bar{\Psi}^{(j)} \gamma_\mu \hat{D}_\nu \Psi^{(j)}-\bar{\Psi}^{(j)} \gamma_\nu \hat{D}_\mu \Psi^{(j)}),
\end{equation}
and the covariant derivative of the conjugate spinor is $\hat{D}_\mu \bar{\Psi}^{(j)}=\partial_\mu\bar{\Psi}^{(j)}+\bar{\Psi}^{(j)}\Gamma_\mu$.

The above action is invariant under a global $U(1)$ transformation $\Psi^{(j)} \rightarrow \Psi^{(j)}e^{i\alpha}$, where $\alpha$ is constant, which implies the conservation of the total particle number. According to Noether’s theorem, we can obtain the Noether charges:
\begin{equation}
\label{Q}
Q=\sum_{j=1}^{2\kappa}Q^{(j)}, \quad  Q^{(j)}=\int_{S} (J^{(j)})^t.
\end{equation}
Here, $S$ is a spacelike hypersurface, $(J^{(j)})^t$ represents the temporal component of the conserved 4-current $(J^{(j)})^{\mu}$. The conserved 4-current can be obtained from the following expression:
\begin{equation}
\label{J}
(J^{(j)})^\mu=\bar{\Psi}^{(j)} \gamma^\mu \Psi^{(j)}.
\end{equation}
ADM mass is also an important global quantity, which can be obtained by integrating the Komar energy density on the spacelike hypersurface $S$:
\begin{equation}
\label{M}
M=\int_{S} T_\mu^\mu-2T_t^t.
\end{equation}

In order to get a spherically symmetric Dirac star solution, we adopt the following metric ansatz:
\begin{equation}
\label{ds2}
d s^2=-e^{2 F_0(r)} d t^2+e^{2 F_1(r)}(d r^2+r^2 d \theta^2+r^2\sin^2\theta d \varphi^2).
\end{equation}
The metric functions $F_0(r)$ and $F_1(r)$  depend only on the radial distance $r$. Additionally, we choose the ansatz for the spinor fields as follows:
%
%
%
\begin{equation}
\label{Dp}
\Psi^{(j)}=\sqrt{\frac{(n+2|m|) ! n !}{(|\kappa|-1) ! |\kappa| !}}\left(\begin{array}{c}
z(r) \\
\bar{z}(r)
\end{array}\right) \otimes\left(\begin{array}{c}
\Theta_1(\theta) \\
\Theta_2(\theta)
\end{array}\right) e^{i(m \varphi-\omega t)}, m=\pm\frac{1}{2}, \pm\frac{3}{2}, ... , \pm(|\kappa|-\frac{1}{2}). \\
\end{equation}
Here, we use a uniform expression, for one of the spinor fields, the specific form is determined by $\kappa$ and $m$. The constant $\omega$ is the angular frequency of the spinor fields, which means that all the spinor fields possess a harmonic time dependence. The part concerning the radial coordinate $r$ is
\begin{equation}
\begin{gathered}
\label{z}
z(r)=ia(r)+b(r),\\
%
\bar{z}(r)=-ia(r)+b(r),
\end{gathered}
\end{equation}
and the part concerning the angular coordinate $\theta$ is
\begin{equation}
\begin{gathered}
\label{Th1}
\Theta_{1}(\theta)=\pm i^{|4m+1|} \left(\sin \frac{\theta}{2}\right)^{|m-\frac{1}{2}|}\left(\cos \frac{\theta}{2}\right)^{|m+\frac{1}{2}|} P_n^{\left(|m-\frac{1}{2}|, |m+\frac{1}{2}|\right)}(\cos \theta),\\
\Theta_{2}(\theta)=\left(\sin \frac{\theta}{2}\right)^{|m+\frac{1}{2}|}\left(\cos \frac{\theta}{2}\right)^{|m-\frac{1}{2}|} P_n^{\left(|m+\frac{1}{2}|, |m-\frac{1}{2}|\right)}(\cos \theta).
\end{gathered}
\end{equation}
%
%
%
%
The real function $a(r)$ and $b(r)$  depend only on the radial distance $r$. In equations (\ref{Th1}), when $\kappa$ is positive, the sign in $\Theta_{1}$ is positive, and when $\kappa$ is negative, the sign in $\Theta_{1}$ is negative. $P_n^{\left(|m+\frac{1}{2}|, |m-\frac{1}{2}|\right)}(\cos \theta)$ and $P_n^{\left(|m-\frac{1}{2}|, |m+\frac{1}{2}|\right)}(\cos \theta)$ are Jacobi polynomials of the $n$-th order, and $n=|\kappa|-|m|-\frac{1}{2}$. The details of the construction process of this ansatz can be found in Appendix B.

By using the above ansatz, we can obtain the specific form of the total energy-momentum tensor:
\begin{equation}
\begin{gathered}
\label{Tto}
T_t^t=-4 e^{-F_0(r)} \omega\left(a(r)^2+b(r)^2\right), \\
T_r^r=4 e^{-F_1(r)}\left(b(r) a^{\prime}(r)-a(r) b^{\prime}(r)\right), \\
T_\theta^\theta=T_{\varphi}^{\varphi}=\frac{4 \kappa e^{-F_1(r)} a(r) b(r)}{r}.
\end{gathered}
\end{equation}
Moreover, the sum of the conserved 4-currents of all spinor fields is independent of $\kappa$:
\begin{equation}
\label{Jsum}
\sum_{j=1}^{|2\kappa|}(J^{(j)})^t=4 e^{-F_0(r)}\left(a(r)^2+b(r)^2\right).
\end{equation}
Except for the temporal component, all other components of the sum of the conserved 4-currents are vanish. 

Equation (\ref{Eequ}) can yield three distinct, independent equations, which are as follows: $E_t^t=0$ , $E_r^r=0$ , $E_\theta^\theta=E_{\varphi}^{\varphi}=0$. By utilizing the equations combinations: $E_\theta^\theta+E_{\varphi}^{\varphi}-E_t^t=0$ and $E_t^t=0$, we can obtain two second-order ordinary differential equations concerning the metric functions:
\begin{equation}
\begin{gathered}
\label{F0F1}
F_0''+F_0'^2+\frac{F_0'}{r}-\frac{1}{2}F_1'^2-\frac{F_1'}{r}-\frac{32\pi \kappa Ge^{F_1}ab}{r}-16\pi \omega Ge^{-F_0+2F_1}(a^2+b^2)=0,\\
F_1''+\frac{1}{2}F_1'^2+\frac{F_1'}{r}+16\pi \omega Ge^{-F_0+2F_1}(a^2+b^2)=0.
\end{gathered}
\end{equation}
And the remaining equation $E_r^r=0$ is employed as a constraint condition to validate the precision of numerical results. Under the ansatz (\ref{Dp}), all spinor fields satisfy the same equations:
\begin{equation}
\begin{aligned}
\label{ab}
& a^{\prime}+\frac{1}{2} a F_0^{\prime}+a F_1^{\prime}+e^{F_1} \mu b-e^{-F_0+F_1} \omega b+\frac{(1+\kappa) a}{r}=0, \\
& b^{\prime}+\frac{1}{2} b F_0^{\prime}+b F_1^{\prime}+e^{F_1} \mu a+e^{-F_0+F_1} \omega a+\frac{(1-\kappa) b}{r}=0.
\end{aligned}
\end{equation}

\section{BOUNDARY CONDITIONS}\label{Sec3}
To numerically solve the above equations, appropriate boundary conditions need to be proposed, which can be determined from the asymptotic behavior of the field functions. For Dirac star solutions, we require them to be asymptotically flat at spatial infinity ($r\rightarrow \infty$). Thus, we need:
\begin{equation}
\label{inf}
F_0(\infty)=F_1(\infty)=a(\infty)=b(\infty)=0.
\end{equation}
At the origin, for the gravitational field, we need:
\begin{equation}
\label{0f}
F_0'(0)=F_1'(0)=0.
\end{equation}
And for the spinor fields, the boundary conditions depend on $\kappa$. 
\begin{equation}
\begin{gathered}
\label{0ab}
\Biggl\{\begin{array}{c}
a(0)=0,\quad b'(0)=0,\quad \kappa=1 \\
a'(0)=0,\quad b(0)=0, \quad \kappa=-1\\
a(0)=0,\quad b(0)=0, \quad |\kappa|\geq2
\end{array}
\end{gathered}
\end{equation}
When $\kappa=\pm 1$, corresponding to Dirac stars with only two spinor fields, the boundary conditions for $a(0)$ and $b(0)$ are different from the case of $|\kappa|\geq2$. The reason for this can be derived from Equation (\ref{ab}), where terms $\displaystyle \frac{(1+\kappa) a}{r}$ and $\displaystyle \frac{(1-\kappa) b}{r}$ lead to the distinction between the cases of $\kappa=\pm 1$ and $|\kappa|\geq2$.

\section{Numerical results}\label{Sec4}
To simplify the equations, we set $\mu=c=G=1$, and $M_{Pl}^2=G^{-1}$ as the Planck mass. In this way, the resulting equations do not involve $\mu$ and $G$. Furthermore, we transform the range of the radial coordinate from $[0,\infty)$ to $[0, 1]$,
\begin{equation}
\label{rtx}
x=\frac{r}{r+1}.
\end{equation}
Our numerical calculations are all based on the finite element method, with 1000 grid points in the integration domain of $0 \leq x \leq 1$. We employ the Newton-Raphson method as the iterative scheme. To ensure the accuracy of the computed results, we require the relative error to be less than $10^{-5}$.

Our work is based on spherically symmetric Dirac stars with positive $\kappa$ and extends them to spherically symmetric Dirac stars with negative $\kappa$. For $\kappa =-1$, distinct boundary conditions are required compared to the case of $\kappa=1$. In our numerical scheme, there are two input parameters: the frequency $\omega$ and the eigenvalue of the Dirac angular operator $\kappa$. It should be noticed that $m$ and $n$ are not treated as parameters since their values depend on $\kappa$. Other physical quantities (such as ADM mass, Noether charge, Komar energy density, etc.) are computed from the numerical solutions. The solution for $\kappa=1$ has already been extensively studied. Here, we will provide the solutions for $\kappa=\pm1,\pm2,\pm3,\pm4,\pm5,\pm6$ and analyze some of their properties.

\begin{figure}[h!]
    \begin{center}
        \includegraphics[height=.27\textheight]{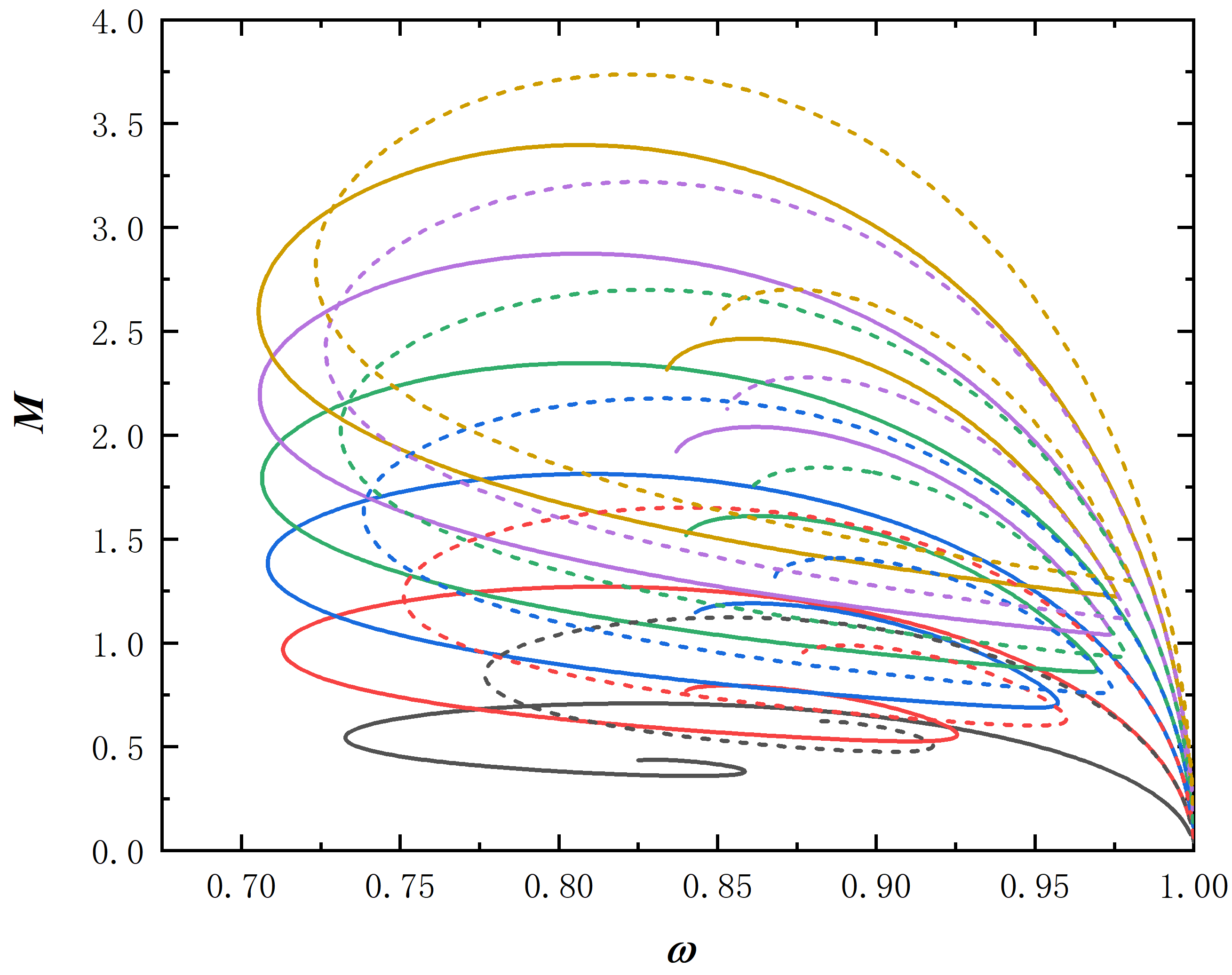}
\includegraphics[height=.27\textheight]{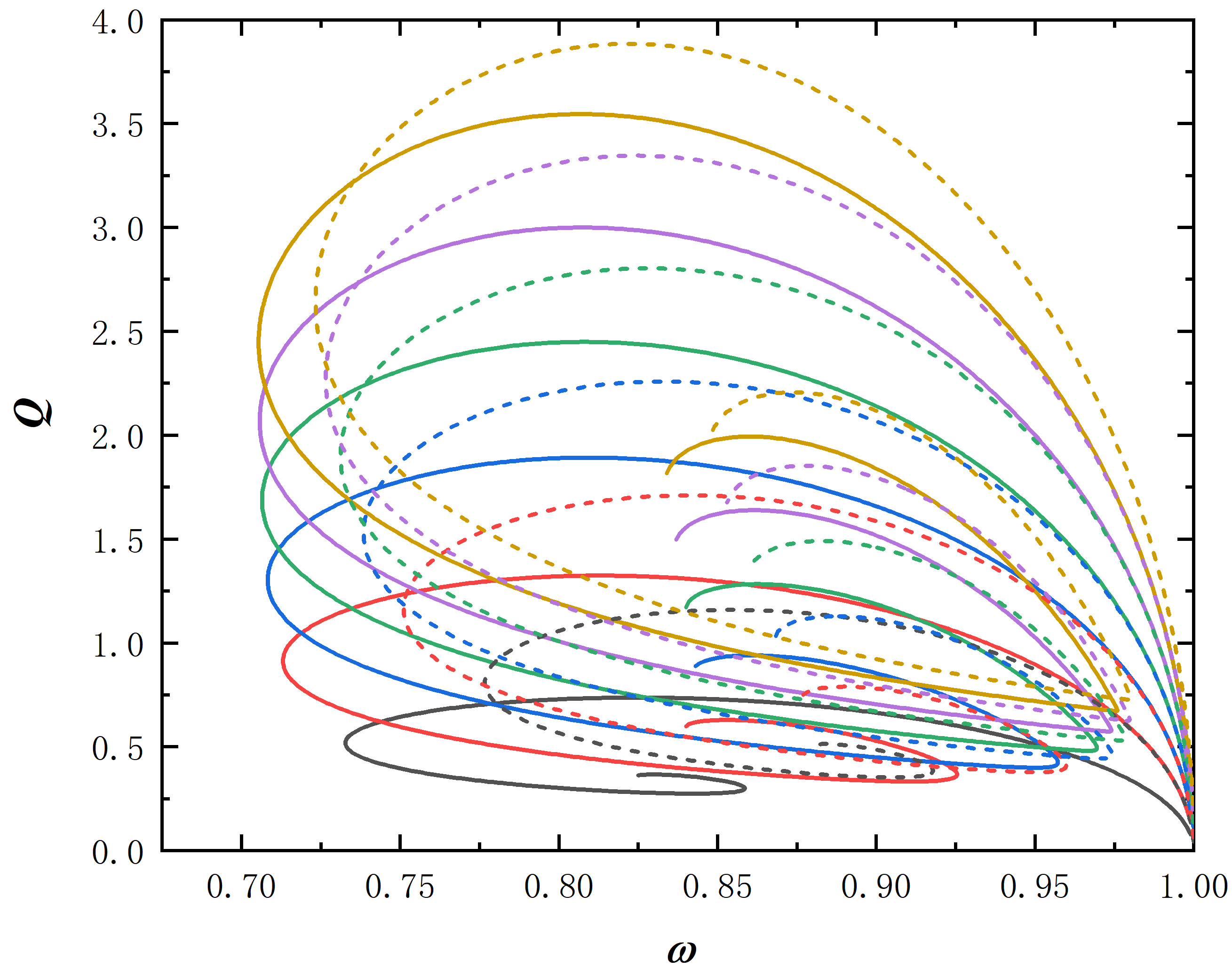}
    \end{center}
    \caption{\small
The ADM mass $M$ (left) and the Noether charge $Q$ (right) as a function of the frequency. The solid lines represent positive $\kappa$ and the dashed lines represent negative $\kappa$. As $|\kappa|$ increases, the color goes from black to yellow. In subsequent figures within this paper, we continue to use these colors to represent different $\kappa$.}
\label{fMQ}
\end{figure}

In Figure \ref{fMQ}, we present the relationship between ADM mass and Noether charge with respect to the frequency for $\kappa=\pm1,\pm2,\pm3,\pm4,\pm5,\pm6$. It can be observed that for different values of $\kappa$, these physical quantities exhibit consistent variations with $\omega$. From the two graphs, we can see that for $M$ and $Q$, starting from the vacuum case, they increase to a maximum value as the frequency decreases, and then decrease as the frequency decreases to a minimum value. We refer to this segment of the curve as the first branch. After the first branch, there is the second branch, where $M$ and $Q$ initially decrease and then increase with increasing frequency, reaching a turning point before entering the third branch. The trend of variation in the third branch is similar to that on the first branch, and the overall curves exhibit a spiral shape. With increasing $|\kappa|$, the curves shifts upward. Furthermore, from Figure \ref{fMQ}, we can observe that the turning points of the second and third branches become sharper as $|\kappa|$ increases.

    	\begin{table}[!t] 
	\centering 
	\begin{tabular}{|c||c|c|c|c|c|c|}
\hline
		&\quad $M_{max} \quad$ & $\quad \omega(M_{max}) \quad$ & $\quad Q_{max} \quad$ & $\quad \omega(Q_{max}) \quad$ & $\quad \omega(M=Q) \quad$ & $\quad \omega_{min} \quad$\\
\hline
		$\kappa=1$ & $0.709$ & $0.825$ & $0.737$ & $0.825$ & $0.743$ & $0.733$ \\
\hline
		$\kappa=-1$ & $1.122$ & $0.854$ & $1.159$ & $0.825$ & $0.786$ & $0.777$ \\
\hline
		$\kappa=2$ & $1.270$ & $0.813$ & $1.324$ & $0.813$ & $0.726$ & $0.713$ \\
\hline
		$\kappa=-2$ & $1.652$ & $0.841$ & $1.710$ & $0.841$ & $0.764$ & $0.751$ \\
\hline
		$\kappa=3$ & $1.813$ & $0.809$ & $1.892$ & $0.809$ & $0.721$ & $0.709$ \\
\hline
		$\kappa=-3$ & $2.177$ & $0.833$ & $2.258$ & $0.833$ & $0.752$ & $0.738$ \\
\hline
		$\kappa=4$ & $2.347$ & $0.808$ & $2.449$ & $0.808$ & $0.719$ & $0.707$ \\
\hline
		$\kappa=-4$ & $2.700$ & $0.828$ & $2.803$ & $0.828$ & $0.745$ & $0.731$ \\
\hline
		$\kappa=5$ & $2.874$ & $0.807$ & $2.999$ & $0.807$ & $0.718$ & $0.706$ \\
\hline
		$\kappa=-5$ & $3.219$ & $0.825$ & $3.345$ & $0.825$ & $0.740$ & $0.727$ \\
\hline
		$\kappa=6$ & $3.396$ & $0.807$ & $3.545$ & $0.807$ & $0.718$ & $0.705$ \\
\hline
		$\kappa=-6$ & $3.736$ & $0.822$ & $3.885$ & $0.822$ & $0.737$ & $0.723$ \\
\hline
	\end{tabular}
 	\caption{Under different values of $\kappa$, the maximum ADM mass $M_{max}$, maximum Noether charge $Q_{max}$ which the Dirac stars can achieve, and the corresponding frequencies $\omega(M_{max})$ and $\omega(q_{max})$. The fifth column provides the frequency $\omega(M=Q)$ at which $M$ and $Q$ are equal. The last column also provides the minimum frequency $\omega_{min}$ of the Dirac stars.}
	\label{tMQ}
\end{table}

Table \ref{tMQ} provides the maximum value of ADM mass $M_{max}$, the maximum value of Noether charge $Q_{max}$, and their corresponding frequencies.  For positive or negative $\kappa$, it can be observed that as $|\kappa|$ increases, the maximum value of ADM mass gradually rises, but the rate of increase diminishes. For two $\kappa$ values with equal absolute value, the maximum ADM mass of solutions with negative $\kappa$ is larger than that of solutions with  positive $\kappa$. The situation is the same for the maximum Noether charge.  Furthermore, for any given $\kappa$ value, the frequency corresponding to the maximum values of ADM mass and Noether charge is identical.  For positive or negative $\kappa$, $\omega(M_{max})$ gradually decreases with an increase in $|\kappa|$, at the same time the rate of decrease becoming smaller. For two $\kappa$ values with equal absolute value, $\omega(M_{max})$ of solutions with negative $\kappa$ is larger than that of solutions with positive $\kappa$. The minimum value of frequency, as well as the frequency when $M$ and $Q$ are equal, both follow the same pattern of variation with respect to $\kappa$.

\begin{figure}[h!]
    \begin{center}
        \includegraphics[height=.32\textheight]{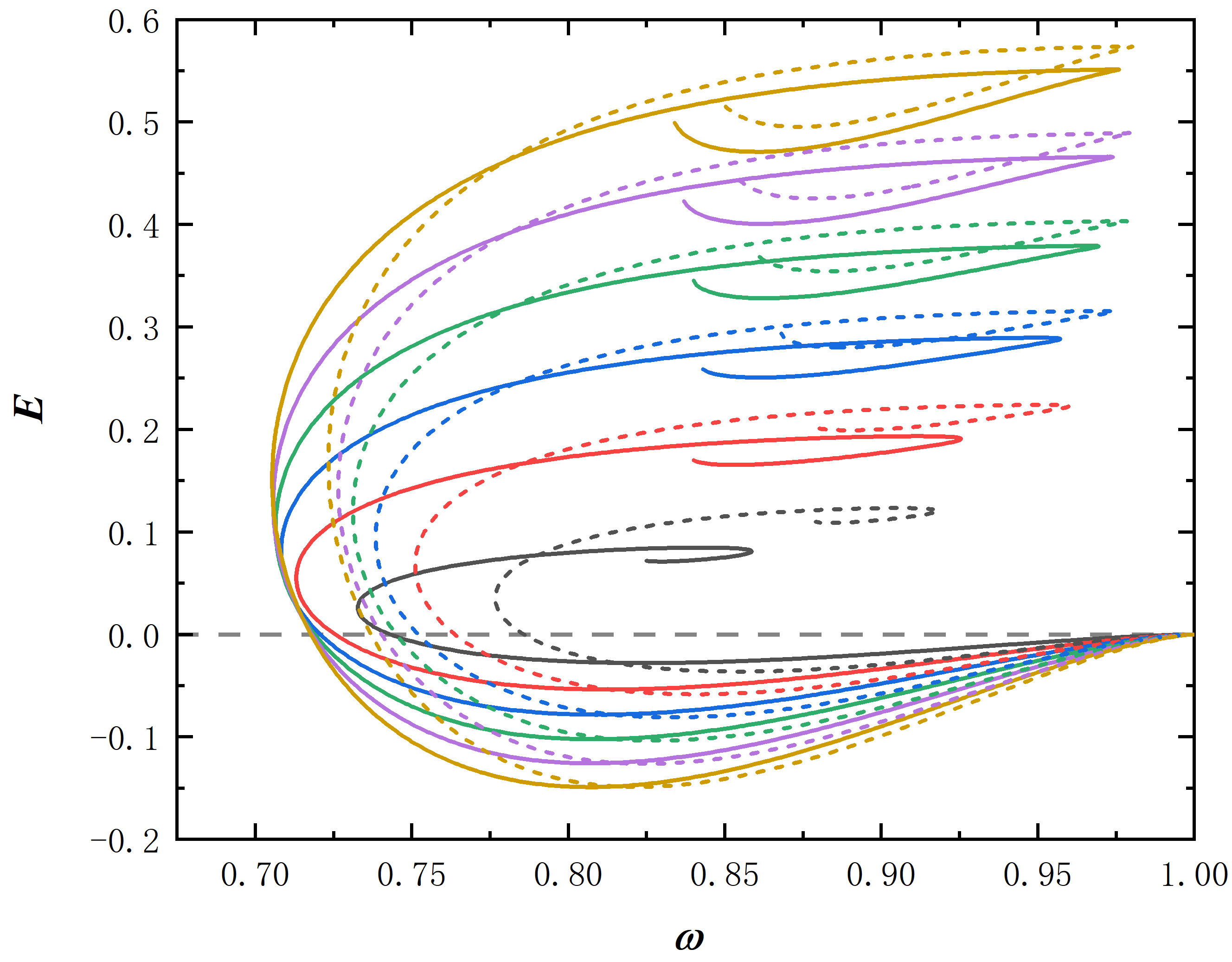}
    \end{center}
    \caption{ The binding energy $E$ as a function of the frequency. The solid lines represent positive $\kappa$ and the dashed lines represent negative $\kappa$. The different colors of lines represent different values of $|\kappa|$.}
\label{fE}
\end{figure}

In addition to ADM mass and Noether charge, the binding energy $E = M - \mu Q$ is also an important physical quantity.  Therefore, in Figure \ref{fE}, we depict the relationship between binding energy and frequency. It can be observed that all curves exhibit the same trend: the binding energy $E$ is zero at the maximum frequency and gradually decreases to a minimum value as the frequency decreases. Then, it increases as the frequency decreases until it enters the second branch. In the second branch, as the frequency increases, the binding energy $E$ increases at first and then decreases. The trend in the third branch is similar with that in the first branch. The curves overall exhibit a helical shape. It can be seen that with an increase in $|\kappa|$, the minimum value of binding energy $E$ becomes smaller, while the maximum value becomes larger. It is generally believed that solutions with positive binding energy are unstable. For all $\kappa$ values, only a portion of the first branch satisfies $E < 0$. From Table \ref{tMQ}, for positive or negative $\kappa$, it can be observed that the frequency corresponding to $E=0$ decreases with an increase in $|\kappa|$, with the rate of decrease becoming smaller. This indicates that the domain of existence for solutions satisfying $E<0$ expands with an increase in $|\kappa|$.

\begin{figure}[h!]
    \begin{center}
        \includegraphics[height=.26\textheight]{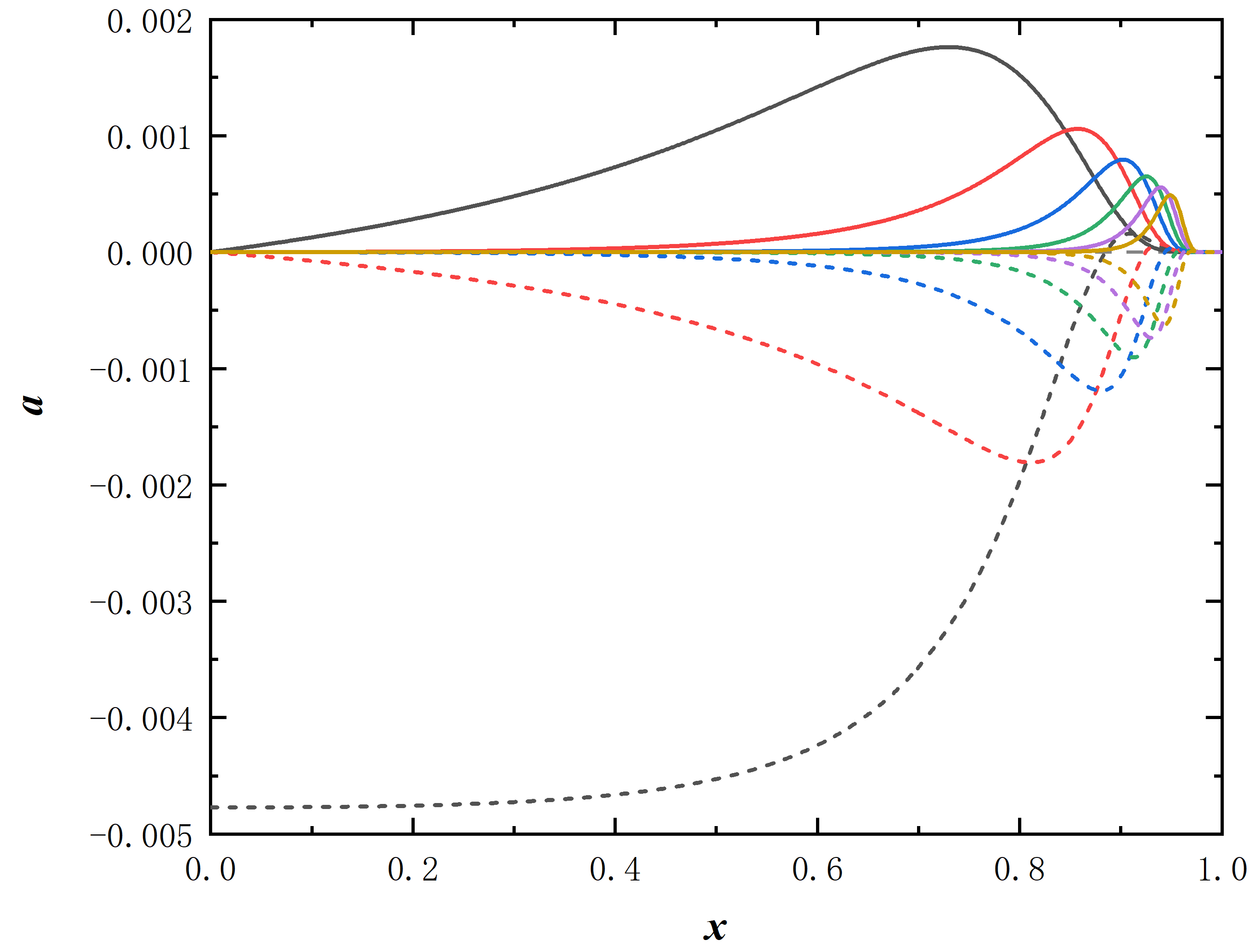}
\includegraphics[height=.26\textheight]{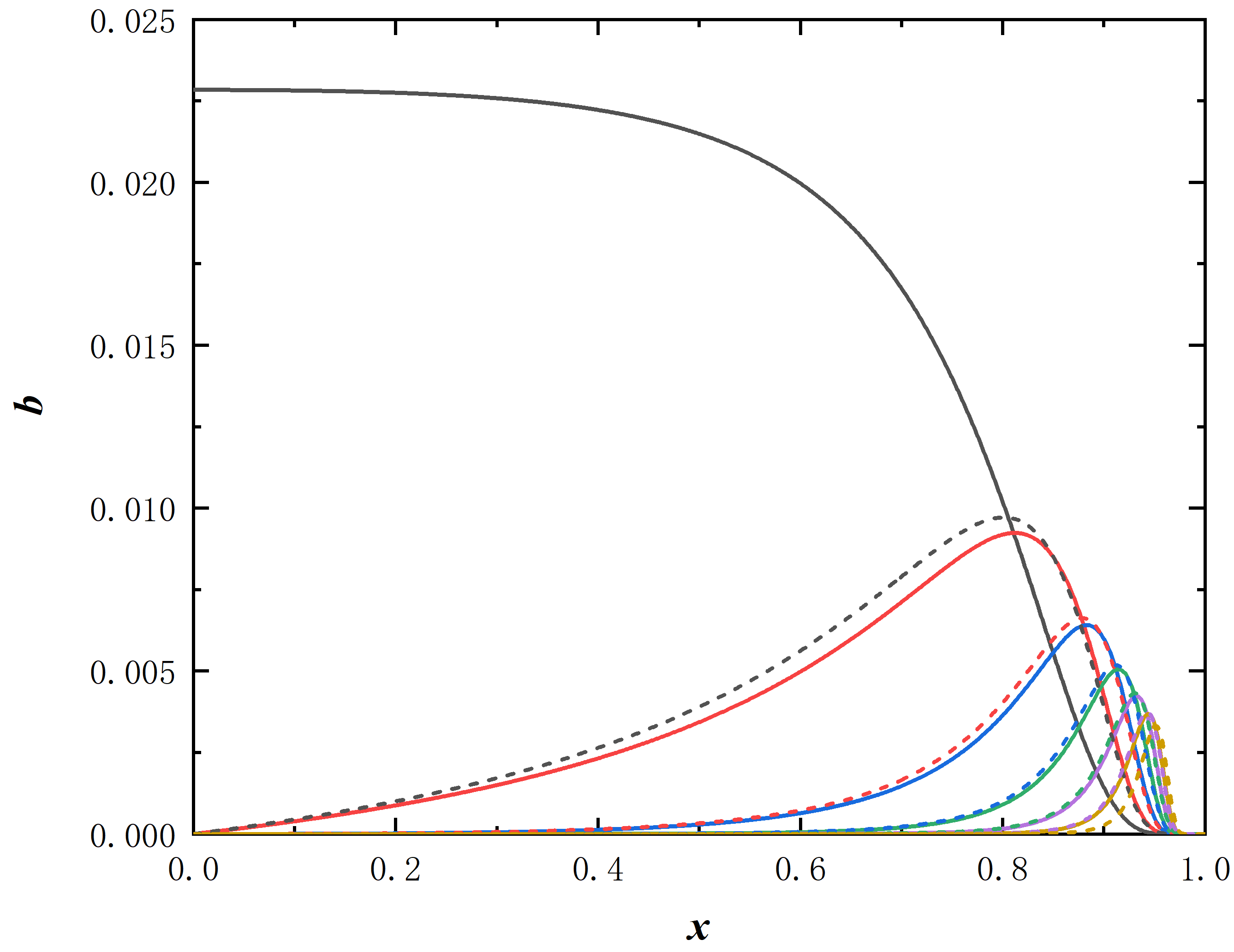}
    \end{center}
    \caption{\small
The radial distribution of the spinor field functions $a$ (left) and $b$ (right). The solid lines represent positive $\kappa$ and the dashed lines represent negative $\kappa$. The different colors of solid lines represent different values of $|\kappa|$.}
\label{fabx}
\end{figure}
\begin{figure}[h!]
    \begin{center}
        \includegraphics[height=.32\textheight]{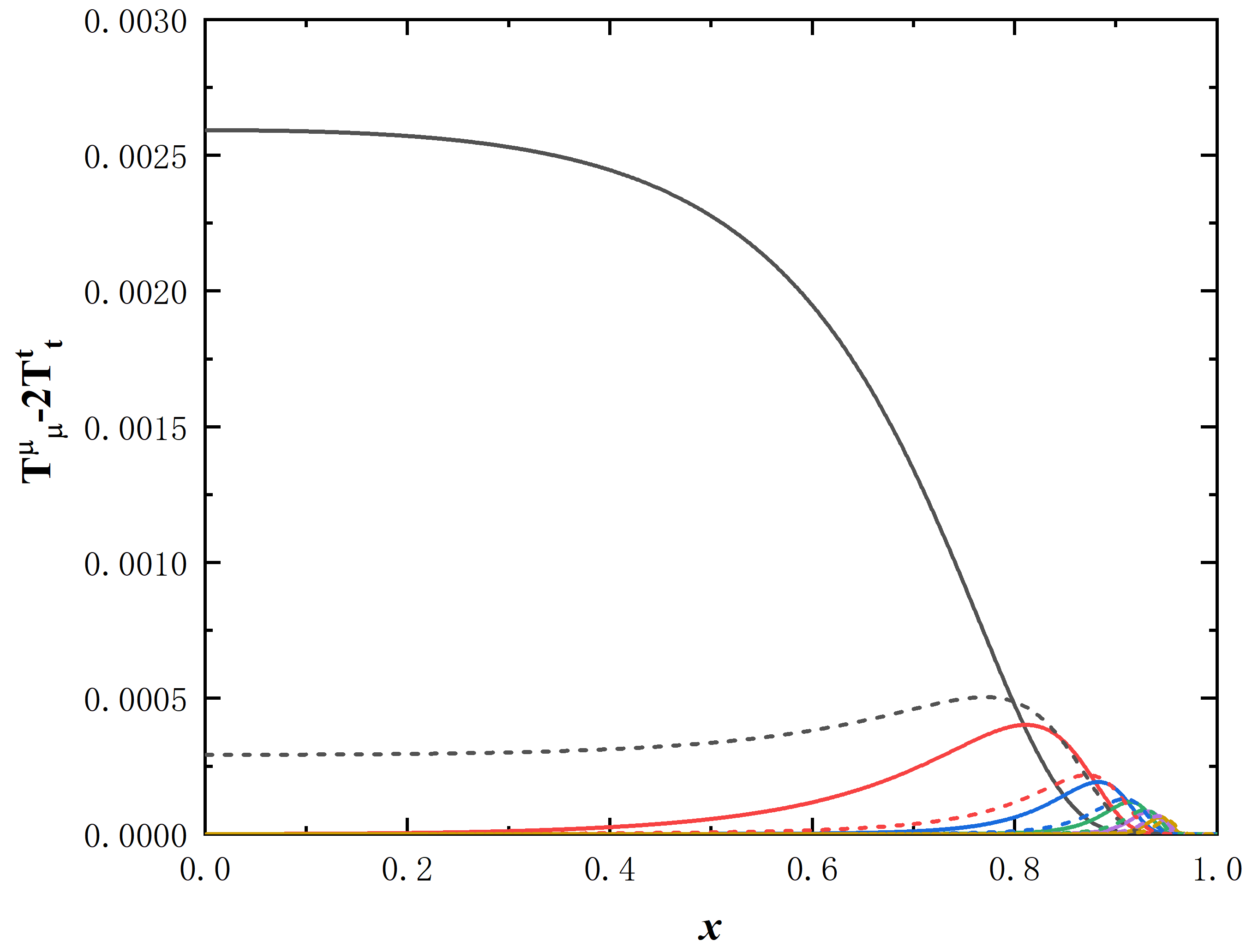}
    \end{center}
    \caption{The radial distribution of the Komar energy density. The solid lines represent positive $\kappa$ and the dashed lines represent negative $\kappa$. The different colors of solid lines represent different values of $\kappa$.}
\label{fTk}
\end{figure}

From equation (\ref{0ab}), we can see that the boundary conditions for the spinor field functions depend on the value of $\kappa$. To visually illustrate this impact, we have plotted the radial distribution of the spinor field functions for different values of $\kappa$ in Figure \ref{fabx}. The field functions correspond to solutions with the frequency  $\omega=0.90$ on the first branch. It can be seen that, except for the case of $\kappa=-1$, the value of $a(0)$ is always zero, and the peak position moves further outward as $|\kappa|$ increases. Furthermore, when $\kappa$ is negative, the function $a$ exhibits a node. For the case of $\kappa = 1$, the field function $b$ has a distribution at the center, while for $\kappa \neq 1$, the field function $b$ is zero at the center. This also leads to a difference in the energy density distribution between the case of $|\kappa| = 1$ and $|\kappa| > 1$. Figure \ref{fTk} displays the radial distributions of  Komar energy density. It can be observed that the energy density distribution is similar to the distribution of the field function $b$. For $\kappa = 1$, there is a peak at the center, while for $|\kappa| > 1$, the central value is $0$, and the distance between the peak and the center increases with increasing $|\kappa|$, and the peak value decreases. However, when $\kappa=-1$, despite the peak not being at the center, the central value is also non-zero. This indicates that for Dirac stars with $|\kappa| > 1$, their shapes resemble more of a shell rather than a solid sphere, unlike the case of $|\kappa| = 1$.

\begin{figure}[h!]
    \begin{center}
        \includegraphics[height=.29\textheight]{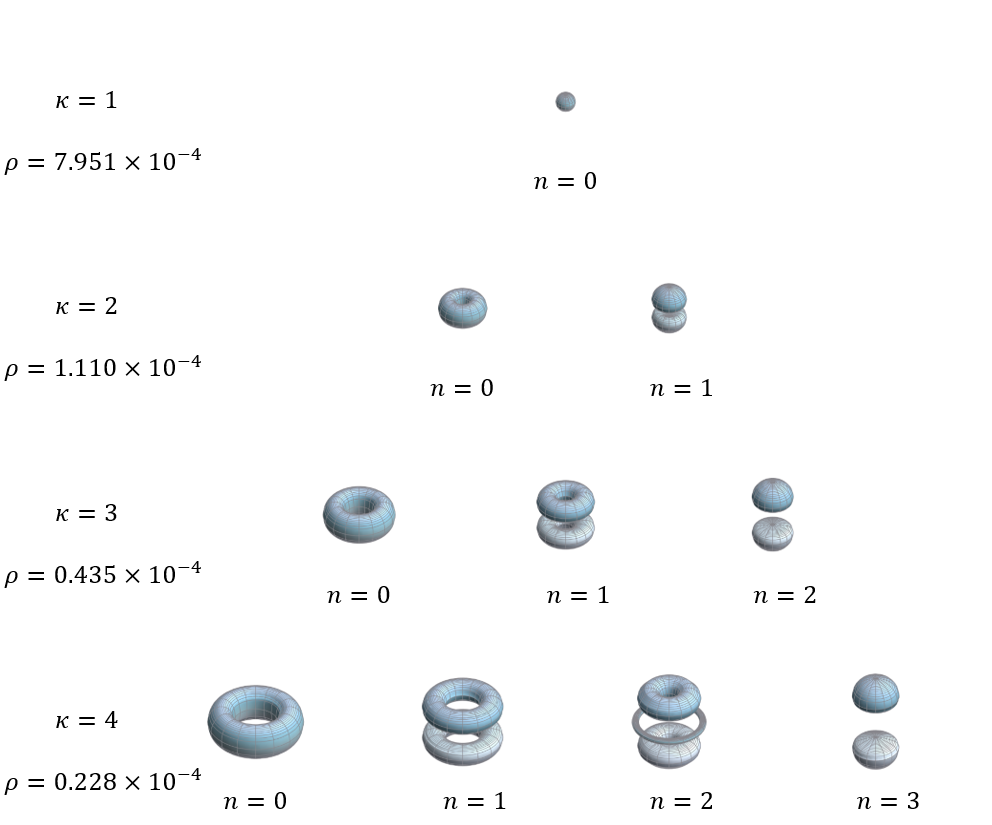}
        \includegraphics[height=.29\textheight]{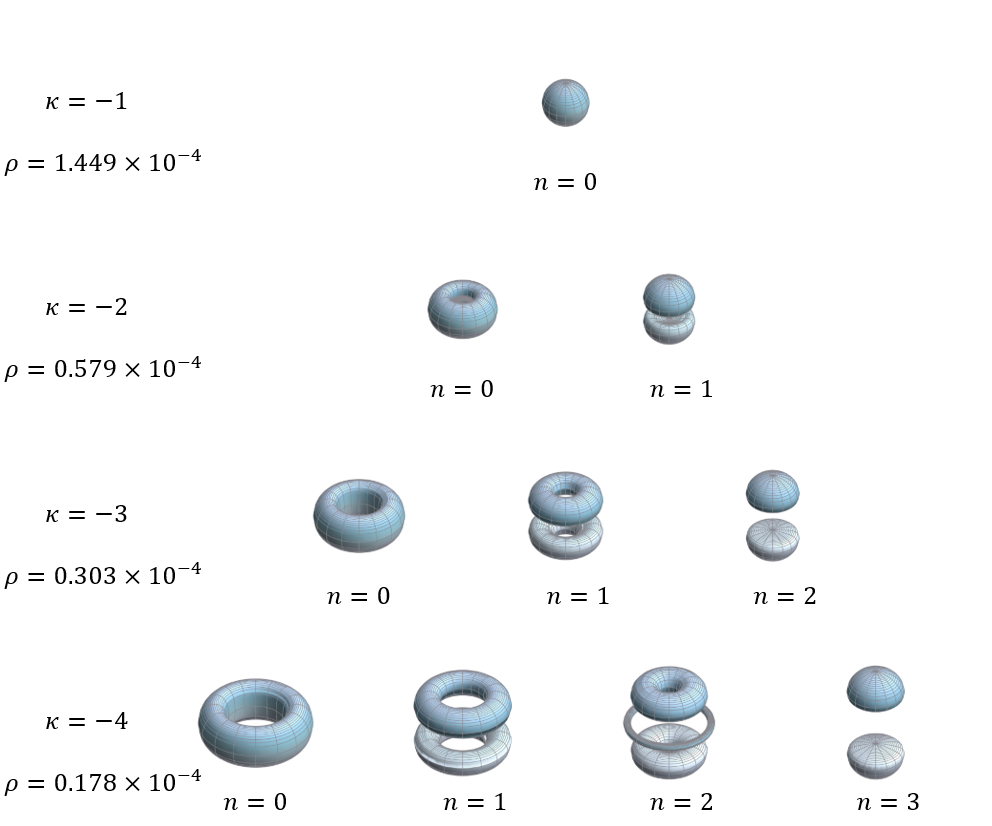}
    \end{center}
    \caption{Surfaces of constant energy density for different values of $\kappa$ and $n$.}
\label{fSE}
\end{figure}

Our model is comprised of $|\kappa|$ pairs of spinor fields, where we consider two spinor fields share the same $n$ value as a pair. Their azimuthal harmonic index $m$ are each other's opposite numbers, and the total angular momentum of this pair of spinor fields vanishes. Although, for a given $\kappa$, all spinor fields satisfy the same Dirac equation, there are still many differences between pairs with different $n$ values. In Figure \ref{fSE}, we display the surfaces of constant energy density for $\kappa=\pm1,\pm2,\pm3,\pm4$ with different values of $n$. Here, the energy density $\rho$ as defined is the time-time component of the energy-momentum tensor, namely: $\rho=-T_{t}^{t}$. Figure \ref{fSE} clearly shows the contribution of each pair of spinor fields in achieving overall spherically symmetric configurations. For the case of $\kappa=\pm1$, $n$ can only take the value of $0$, which means there is only one pair of spinor fields, and it exhibits clear spherically symmetric behavior. As for $\kappa=\pm2$, the pair with $n=0$ forms a torus-shaped structure, while the pair with $n=1$ exhibits two ellipsoidal shapes symmetric about the equatorial plane. For $\kappa=\pm3$, the pair with $n=0$ remains toroidal, while the pair with $n=1$ forms two tori symmetric about the equatorial plane, and the pair with $n=2$ shows two ellipsoidal shapes symmetric about the equatorial plane. In the case of $\kappa=\pm4$, in addition to the three types mentioned above, there is an additional structure with three rings, where the ring on the equatorial plane is narrower than others. This corresponds to $n=2$. It can be observed that as $|\kappa|$ increases, the distribution regions along the axis of symmetry become narrower for each component. Furthermore, it is noteworthy that only when $|\kappa| - n=1$ (i.e., $|m|=1/2$), the spinor fields can be distributed along the axis of symmetry. The $\ell$-boson stars \cite{Alcubierre:2018ahf} which have $2\ell+1$ scalar fields, can be regarded as a combination of multipolar boson stars \cite{Herdeiro:2020kvf,Sanchis-Gual:2021edp}. The $\kappa$-Dirac stars model have $|2\kappa|$ spinor fields. Each pair of spinor fields also exhibits an energy distribution similar to that of multipolar boson stars.

\section{Conclusion}\label{sec5}

In this article, we construct Dirac stars formed by $|\kappa|$ pairs of spinor fields. These spinor fields satisfy the same Dirac equation, and different solutions can be labeled by the value of $\kappa$. We present solutions for $\kappa=\pm1,\pm2,\pm3,\pm4,\pm5,\pm6$ and analyzed their properties.

Firstly, we provide the relationship between ADM mass $M$ and Noether charge $Q$ with respect to frequency $\omega$. They exhibit a spiral-like shape in the $M-\omega$ and $Q-\omega$ plots. As $|\kappa|$ increases, both $M$ and $Q$ increase. We also present a table comparing the maximum values of $M$ and $Q$ along with their corresponding frequencies. For all $\kappa$, $M$ and $Q$ reach their maximum values at the same frequency. For positive or negative $\kappa$, with an increase in $|\kappa|$, the frequency corresponding to the maximum values of $M$ and $Q$ decreases. The frequency at which $M$ and $Q$ are equal also decreases gradually. Additionally, we provide the minimum value of frequency, which decreases with an increase in $|\kappa|$.

To assess the stability-related properties of the Dirac stars, we provide the binding energy as a function of frequency. We observe that for any $\kappa$, only a portion of the first branch satisfies $E < 0$. This indicates that, from the perspective of binding energy, only this portion is stable. On the other hand, as $|\kappa|$ continues to increase, the minimum value of $E$ becomes smaller, and the frequency at $E=0$ decreases. This suggests that the domain of existence for solutions satisfying $E < 0$ expands, and larger $|\kappa|$ values may possess better stability.

We also provid the radial distribution of the Komar energy density for different $\kappa$. It is evident that for $|\kappa|>1$, the spinor fields are not distributed at the center, unlike the case of $|\kappa|=1$. This indicates that the Dirac stars form a shell-like configuration for $|\kappa|>1$ while being spherical for $|\kappa|=1$. Moreover, at the same frequency, the shell becomes larger with increasing $|\kappa|$. In Figure \ref{fSE}, we showe the surfaces of constant energy density for $\kappa=\pm1,\pm2,\pm3,\pm4$ distinguished by the value of n, which provide a clear understanding of how these $|\kappa|$ pairs of spinor fields achieve overall spherically symmetric configurations.

We observe that as $|\kappa|$ increased, the maximum value of $M$, the maximum value of $Q$, and the minimum value of frequency all exhibit a trend of approaching a limit. Therefore, exploring this system in the limit $|\kappa|\rightarrow \infty$ to quest any special properties that might emerge, akin to what is done with $\ell$-boson stars \cite{Alcubierre:2021psa}, can be intriguing. Additionally, this system can be extended to higher dimensions; however, as the dimension increases, more spinor fields may be needed to form a spherically symmetric spacetime.

\section*{ACKNOWLEDGEMENTS}\label{ack}

This work is supported by the National Key Research and Development Program of China (Grant No. 2020YFC2201503) and the National Natural Science Foundation of China (Grants No. 12275110 and No. 12047501). Parts of computations were performed on the shared memory system at the Institute of computational physics and complex systems at Lanzhou University. 	

\section*{Appendix A}\label{apA}
Within this appendix, we shall elucidate the procedure for obtaining the Dirac equation in the context of curved spacetime.  From the metric in equation (\ref{ds2}), we naturally obtain the vierbein :
\begin{equation}
\label{e}
e_{\mu}^{a}=\left(\begin{array}{cccc}
e^{F_0(r)} \quad & 0 \quad & 0 \quad & 0 \\
0 \quad & e^{F_1(r)} \quad & 0 \quad & 0 \\
0 \quad & 0 \quad & e^{F_1(r)}r \quad & 0 \\
0 \quad & 0 \quad & 0 \quad & e^{F_1(r)}r\sin\theta 
\end{array}\right),
\end{equation}
which satisfies
\begin{equation}
\label{ee}
g_{\mu\nu}=e_{\mu}^{a}e_{\nu a}, \quad \eta_{ab}=e_{a}^{\mu}e_{b\mu}.
\end{equation}
Here, $\eta_{ab}=\text{diag}(-1,1,1,1)$ is the Minkowski metric. Imposing vierbein compatibility 
\begin{equation}
\label{pte}
\partial_{\mu} e_{\nu}^{a}+\omega_{\mu \  b}^{\  a} e_{\nu}^{b} - \Gamma^{\lambda}_{\ \nu\mu} e_{\lambda}^{a}=0,
\end{equation}
where $\Gamma^{\lambda}_{\ \nu\mu}$ is the affine connection. This equation can lead to the spin connection
\begin{equation}
\label{pte}
\omega_{\mu \  b}^{\  a}=e_{\nu}^{a}e_{b}^{\lambda}\Gamma^{\nu}_{\ \mu\lambda}-e_{b}^{\lambda}\partial_{\mu} e_{\lambda}^{a}.
\end{equation}
And we can get the spin connection matrices
\begin{equation}
\label{Gamma}
\Gamma_{\mu}= -\frac{1}{4}\omega_{\mu ab}\hat{\gamma}^{a}\hat{\gamma}^{b}.
\end{equation}
In the flat spacetime, the gamma matrices $\hat{\gamma}^{a}$ which we choosed are as follows:
\begin{equation}
\label{gam}
\hat{\gamma}^{0}=i\sigma_1\otimes \sigma_0 \quad \hat{\gamma}^{1}=\sigma_2\otimes \sigma_0\quad \hat{\gamma}^{2}=\sigma_3\otimes \sigma_1\quad \hat{\gamma}^{3}=\sigma_3\otimes \sigma_2,
\end{equation}
where the symbol $\otimes$ stands for the direct product and
\begin{equation}
\label{pau}
\sigma^{0}=\left(\begin{array}{cc}
1 \quad & 0 \\
0 \quad & 1  
\end{array}\right) \quad \sigma^{1}=\left(\begin{array}{cc}
0 \quad & 1 \\
1 \quad & 0  
\end{array}\right) \quad \sigma^{2}=\left(\begin{array}{cc}
0 \quad & -i \\
i \quad & 0 
\end{array}\right) \quad \sigma^{3}=\left(\begin{array}{cc}
1 \quad & 0 \\
0 \quad & -1  
\end{array}\right),
\end{equation}
are unit matrix and Pauli matrices. We can get the gamma matrices in curve spacetime:
\begin{equation}
\label{egam}
\gamma^{\mu}=e_{a}^{\mu}\hat{\gamma}^{a}.
\end{equation}
It is readily ascertainable that the gamma matrices $\gamma^{\mu}$ and $\hat{\gamma}^{a}$ satisfy the anti-commutation relations:
\begin{equation}
\label{gamgam}
\{\gamma^{\mu},\gamma^{\nu}\}= 2g^{\mu\nu}I_{4}, \quad \{\hat{\gamma}^{a},\hat{\gamma}^{b}\}= 2\eta^{ab}I_{4},
\end{equation}
where $\{A,B\}=AB+BA$.

\section*{Appendix B}\label{apB}
In order to attain spherically symmetric solutions for the Einstein-Dirac system, the requisite conditions that the spinor field must initially satisfy are as follows: the radial component of the ansatz can be separated from the angular component, and the angular part is an eigenfunction of the Dirac angular operator. The eigenvalue problem of the Dirac operator on the sphere was studied as early as 1966 by Newman and Penrose \cite{Leith:2021urf}. Subsequent research has provided specific forms of the eigenfunctions \cite{Abrikosov:2002jr}. The Dirac angular operator is the part of the Dirac operator  that depends only on the angles in spherically symmetric curved spacetime as:
\begin{equation}
\label{K}
\hat{K}=-i\sigma_{1}\left(\partial_{\theta}+\frac{\cot\theta}{2}\right)-i\sigma_{2}\frac{\partial_{\varphi}}{\sin\theta}.
\end{equation}
We assume that the Dirac four-component spinor $\Psi$ can be separated into a direct product form:
\begin{equation}
\label{KPsi}
\Psi=e^{-i\omega t}R(r)\otimes\Theta(\theta,\varphi),
\end{equation}
where $R(r)$ and $\Theta(\theta,\varphi)$ both have tow components. It can be seen that if $\Theta(\theta,\varphi)$ is an eigenfunction of the Dirac angular operator $\hat{K}$, the Dirac equation reduces to a set of ordinary differential equations depending only on $r$. So, we have
\begin{equation}
\label{Kth}
\hat{K}\Theta(\theta,\varphi)=\kappa\Theta(\theta,\varphi),
\end{equation}
where $\kappa$ is the eigenvalue.

Next, we need to solve the eigenvalue equation of $\hat{K}$. This first-order partial differential equation allows us to separate variables for $\Theta(\theta,\varphi)$ as:
\begin{equation}
\begin{gathered}
\label{mth}
\Theta(\theta,\varphi)=\left(\begin{array}{c}
\Theta_{1}(\theta) \\
\Theta_{2}(\theta)
\end{array}\right)e^{im\varphi},
\end{gathered}
\end{equation}
where $m$ is a half-integer and is regarded as the projection of angular momentum along the polar axis. The eigenvalue equation can be written as:
\begin{equation}
\begin{gathered}
\label{Kmth}
-i\left(\partial_{\theta}+\frac{\cot\theta}{2}+\frac{m}{\sin\theta}\right)\Theta_{2}(\theta)=\kappa\Theta_{1}(\theta),\\
-i\left(\partial_{\theta}+\frac{\cot\theta}{2}-\frac{m}{\sin\theta}\right)\Theta_{1}(\theta)=\kappa\Theta_{2}(\theta).
\end{gathered}
\end{equation}
Applying $\hat{K}$ once again to both sides of the equation (\ref{Kth}), we obtain a set of separate second-order equations for the two components:
\begin{equation}
\begin{gathered}
\label{K2th}
\left(-\frac{1}{\sin\theta}\partial_{\theta}\sin\theta\partial_{\theta}+\frac{m^{2}}{\sin^{2}\theta}\mp\frac{m\cos\theta}{\sin^{2}\theta}+\frac{1}{4}+\frac{1}{4\sin^{2}\theta}\right)\left(\begin{array}{c}
\Theta_{1}(\theta) \\
\Theta_{2}(\theta)
\end{array}\right)=\kappa^{2}\left(\begin{array}{c}
\Theta_{1}(\theta) \\
\Theta_{2}(\theta)
\end{array}\right).
\end{gathered}
\end{equation}
This is a generalized hypergeometric equation, and by studying hypergeometric equations, we can obtain the solutions for these two components:
\begin{equation}
\begin{gathered}
\label{Thab}
\Theta_{1}(\theta)=\pm i^{|4m+1|} \left(\sin \frac{\theta}{2}\right)^{|m-\frac{1}{2}|}\left(\cos \frac{\theta}{2}\right)^{|m+\frac{1}{2}|} P_n^{\left(|m-\frac{1}{2}|, |m+\frac{1}{2}|\right)}(\cos \theta),\\
\Theta_{2}(\theta)=\left(\sin \frac{\theta}{2}\right)^{|m+\frac{1}{2}|}\left(\cos \frac{\theta}{2}\right)^{|m-\frac{1}{2}|} P_n^{\left(|m+\frac{1}{2}|, |m-\frac{1}{2}|\right)}(\cos \theta),
\end{gathered}
\end{equation}
where $P_n^{\left(|m+\frac{1}{2}|, |m-\frac{1}{2}|\right)}(\cos \theta)$ denotes the Jacobi polynomial of order $n$. When $\kappa$ is positive, the sign in $\Theta_{1}$ is positive, and when $\kappa$ is negative, the sign in $\Theta_{1}$ is negative. The eigenvalue $\kappa$ satisfies the relationship $|\kappa|= |m|+n+\frac{1}{2}$. This signifies that for any given $\kappa$, there exist $|2\kappa|$ eigenfunctions $\Theta$ satisfying the eigenvalue equation (\ref{Kth}).

For a single spinor field, $T_{t\varphi}$ is non-zero due to its angular momentum, preventing the construction of a static spherically symmetric spacetime. To obtain a solution for a static spherically symmetric spacetime, it is necessary to simultaneously consider a pair of spinor fields with equal magnitudes and opposite directions of angular momentum, such that the $T_{t\varphi}$ becomes zero. Specifically, when considering the spinor field with $m$, we also need to consider the spinor field with $-m$.

In addition, to ensure the vanishing of $T_{tr}$, it is requisite for the two components $R_1$ and $R_2$ of $R(r)$ to satisfy the following conditions: $|R_{1}|^{2} = |R_{2}|^{2}$. For the sake of simplicity,  we choose:
\begin{equation}
\begin{gathered}
\label{Thab}
R_1(r)=z(r)=ia(r)+b(r),\\
R_2(r)=\bar{z}(r)=-ia(r)+b(r).
\end{gathered}
\end{equation}
Here, $a(r)$ and $b(r)$ are real functions.

By satisfying the above requirements, we obtain a diagonal total energy-momentum tensor. However, for $|m| > 1/2$, if we only consider the combination of spinor fields with $m$ and $-m$, $T_{\mu}^{\nu}$ is not solely dependent on $r$. This leads to the incapability of the Einstein equation to be satisfied.  This elucidates that for $|m| >  1/2$, relying solely on the combination of two spinor fields cannot yield solutions for static spherically symmetric spacetimes. We find that, for any $m$, it requires combinations of $2|m| + 1$ spinor fields to construct a static spherically symmetric spacetime. These spinor fields have the same eigenvalue $\kappa$ for their angular part $\Theta$ (which satisfies the same Dirac equation), and they are combined with specific coefficient factors $\displaystyle \sqrt{\frac{(n+|2m|) ! n !}{(|\kappa|-1) ! |\kappa| !}}$. The final form of the spinor fields used is equation (\ref{Dp}).

\providecommand{\href}[2]{#2}\begingroup\raggedright
\endgroup

\end{CJK*}
\end{document}